\input{aipcheck}

\documentclass[
    ,final         
  ]
  {aipproc}

\layoutstyle{8x11single}

\begin{document}

\title{Impact picture for near-forward elastic scattering up to LHC energies}


\author{Jacques Soffer}{
  address={Physics Department, Temple University, 1835 N, 12 Street, Philadelphia, PA 19122-6082, USA}
}

\author{Claude Bourrely}{
  address={Aix-Marseille Universit\'e, D\'epartement de Physique, Facult\'e des Sciences, site de Luminy, 13288 Marseille, Cedex 09, France}
}

\author{Tai Tsun Wu}{
  address={Gordon McKay Laboratory, Harvard University, Cambridge, MA 02138, USA }
,altaddress={Theoretical Physics Division, CERN, 1211 Geneva 23, Switzerland} 
}

\begin{abstract}
 We will recall the main feaatures of an accurate phenomenological model to describe successfully near-forward elastic scattering in a wide energy range, including ISR, SPS and Tevatron colliders. A large step in energy domain is accomplished with the LHC collider, presently running, giving the opportunity to confront the new data with the predictions of our theoretical approach.
\end{abstract}

\maketitle


\section{Introduction}
The measurements of high energy  $\bar p p~\mbox{and}~p p$ elastic at ISR, SPS, and Tevatron colliders
have provided usefull informations on the behavior of the scattering
amplitude, in particular, on the nature of the Pomeron. A large step in energy
domain is accomplished with the LHC collider presently running, giving a unique
opportunity to improve our knowledge on the asymptotic regime of the scattering
amplitude and to verify the validity of our approach. We will first
recall the basic ingredients of the BSW amplitude and its essential features. We
will also mention the success of its predictions so far in the energy range below the LHC energy, for the total cross section
$\sigma_{tot}(s)$, the ratio of the real to imaginary parts of the forward amplitude $\rho(s)$ and the differential cross section 
${d\sigma (s,t)/dt}$. Our predictions at LHC will be shown
and compared with the first experimental results and we will recall why its is so important to measure $\rho$ at LHC

\section{Main features of the BSW model}
The BSW model was first proposed, in 1978 \cite{bsw}, to describe the experimental data on elastic $p p$ and $\bar{p} p$, taken at the relatively low energies available
to experiments, forty years ago or so. Some more complete analysis were done later \cite{bsw1,bsw2,bsw3},  showing very successful theoretical predictions for these processes. Since a new energy domain is now accessible with the LHC collider at CERN, it is a good time to recall the main features of the BSW model and to check its validity. The spin-independent elastic scattering amplitude is given by
\begin{equation}
  \label{eq:one}
  a(s,t) = \frac{is}{2\pi}\int e^{-i\mathbf{q}\cdot\mathbf{b}} (1 - 
e^{-\Omega_0(s,\mathbf{b})})  d\mathbf{b} \ ,
\end{equation}
where $\mathbf q$ is the momentum transfer ($t={-\bf q}^2$) and 
$\Omega_0(s,\mathbf{b})$ is the opaqueness at impact parameter 
$\mathbf b$ and at a given energy $s$, the square of the center-of-mass energy. We take the simple form
\begin{equation}
\label{eq:omega}
\Omega_0(s,\mathbf{b}) = S_0(s)F(\mathbf{b}^2)+ R_0(s,\mathbf{b}) \, ,
\end{equation}
the first term is associated with the "Pomeron" exchange, which generates 
the diffractive component of the scattering and the second term is 
the Regge background which is negligible at high energy.
The function $S_0(s)$ is given by the complex symmetric expression, obtained from the high energy behavior
of quantum field theory \cite{chengWu}
\begin{equation}
  \label{eq:Sdef}
  S_0(s)= \frac{s^c}{(\ln s)^{c'}} + \frac{u^c}{(\ln u)^{c'}}~,
\end{equation}
with $s$ and $u$ in units of $\mbox{GeV}^2$, where $u$ is the
third Mandelstam variable. In Eq. (\ref{eq:Sdef}), $c$ and $c'$ are two 
dimensionless constants given above \footnote{In the Abelian case one finds $c' = 3/2$ and it was
conjectured that in Yang-Mills non-Abelian gauge theory
one would get $c' = 3/4$ (T.T. Wu private communication).}
in Table 1. That they
are constants implies that the Pomeron is a fixed Regge cut rather than a
Regge pole. For the asymptotic behavior at high energy and modest momentum
transfers, we have to a good approximation
\begin{equation}
  \label{eq:uAp}
  \ln u = \ln s -i\pi~,
\end{equation}
so that
\begin{equation}
  \label{eq:S2}
    S_0(s)= \frac{s^c}{(\ln s)^{c'}} + 
\frac{s^ce^{-i\pi c}}{(\ln s-i\pi)^{c'}}~.
\end{equation}
The choice one makes for $F(\mathbf{b}^2)$ is essential and we take the Bessel transform of
\begin{equation}
  \label{eq:FtildDef}
  \tilde{F}(t) = f[G(t)]^2\frac{a^2+t}{a^2-t}~,
\end{equation}
where $G(t)$ stands for the proton "` nuclear form factor"', parametrized similarly to the electromagnetic
form factor, with two poles
\begin{equation}
  \label{eq:Gdef}
  G(t)=\frac{1}{(1- t/m_1^2)(1- t /m_2^2)}~.
\end{equation}
The remaining four parameters of the model, $f$, $a$, $m_1$ $\mbox{and}$ $m_2$, are given in Table 1. It is interesting to observe that the BSW parameters which have been determined in 1979 and 1984, exhibit a remarkable stability.\\
We define the ratio of the real to imaginary parts of the forward amplitude, mentioned earlier in the introduction, 
$\rho(s) = \frac{\mbox{Re}~a(s, t=0)}{\mbox{Im}~a(s, t=0)}$,
 the total cross section 
$\sigma_{tot} (s) = (4\pi/s)\mbox{Im}~a(s, t=0)$, 
the differential cross section
${d\sigma (s,t)/dt} = \frac{\pi}{s^2}|a(s,t)|^2$, 
and the integrated elastic cross section 
$\sigma_{el} (s) = \int dt \frac{d\sigma (s,t)} { dt}$. One important feature of the BSW model is, as a consequence of Eq. (\ref{eq:S2}), the fact that the phase of the amplitude is built in. Therefore real and imaginary parts of the amplitude cannot be chosen independently.\\
In the next section we will recall some of the early successes of our approach at the CERN $\bar {p} p$ collider
and at the FNAL Tevatron and the last section will be devoted to a discussion of the sitution at the Large Hadron Collider.

\begin{table}
\begin{tabular}{|c|c|c|}
\hline
Year & 1979 & 1984\\
\hline\raisebox{0pt}[12pt][6pt]{$c$}&0.151    & 0.167     \\[4pt]
\hline\raisebox{0pt}[12pt][6pt]{$c'$}&0.756    & 0.748     \\[4pt]
\hline\raisebox{0pt}[12pt][6pt]{$m_1$}&0.619    & 0.586     \\[4pt]
\hline\raisebox{0pt}[12pt][6pt]{$m_2$}&1.587    & 1.704     \\[4pt]
\hline\raisebox{0pt}[12pt][6pt]{$f$}  &8.125    & 7.115     \\[4pt]
\hline\raisebox{0pt}[12pt][6pt]{a}    &2.257    & 1.953     \\[4pt]
\hline
\end{tabular}
\caption{ Pomeron fitted parameters for $pp (\bar p p)$ comparing the 1979 \cite{bsw} and 1984 \cite{bsw1} solutions}
\label{table:a}
\end{table} 

\section{Pre LHC era successes}
At the FNAL-Tevatron, the E710 experiment running at $\sqrt{s}$=1.8TeV, has obtained $\sigma_{tot}=72.8 \pm 3.1$ mb and $\sigma_{el}/\sigma_{tot}= 0.23 \pm 0.012$ \cite{amos}, whereas the BSW predictions are 74.8 mb and 0.230 respectively (See Fig. 1). They were also able to extract the following $\rho$ value, $\rho = 0.140 \pm 0.069$ \cite{amos2}. This important measurement is in agreement with the BSW prediction, but has unfortunately little significance because of its lack of precision. These data are reported in Fig. 1 together with the results of the CDF experiment at two different Tevatron energies $\sqrt{s}$=1.8TeV and $\sqrt{s}$=546GeV \cite{cdf}
and the results of UA(4) at the CERN $\bar p p$ collider at $\sqrt{s}$=541GeV \cite{augier2}. At $\sqrt{s}$=1.8TeV CDF found $\sigma_{tot}=80.03 \pm 2.24$ mb and $\sigma_{el}/\sigma_{tot}= 0.246 \pm 0.004$, at variance with the E710 results. However at $\sqrt{s}$=546GeV, the CDF results $\sigma_{tot}=61.26 \pm 0.93$ mb and $\sigma_{el}/\sigma_{tot}=0.210 \pm 0.002$ agree well with those of UA(4), $\sigma_{tot}=63.0 \pm 2.1$ mb and $\sigma_{el}/\sigma_{tot}=0.208 \pm 0.007$. The UA(4) experiment has obtained a very precise value for
the parameter $\rho$, $\rho=0.135 \pm 0.015$, from the measurement of $d\sigma/dt$ in the Coulomb-nuclear interference region \cite{augier}, as shown
in Fig. 2. One notices the rapid rise of the cross section in the very low $t$ region and the remarquable agreement with the BSW prediction. 
The BSW model predicts the correct $\rho(s)$ which appears to have a flat energy dependence in the high energy region  and for $s \to \infty$, one expects $\rho(s) \to 0$. Another specific feature of the BSW model is the fact that it incorporates the theory of expanding protons \cite{chengWu}, with the physical consequence that the ratio $\sigma_{el}/\sigma_{tot}$ increase with energy. This is precisely in agreement with the data and when $s \to \infty$ one expects $\sigma_{el}/\sigma_{tot} \to 1/2$, which is the black disk limit.

Finally we show in Fig. 2 the $t$-dependence of the elastic cross section measured by the E710 experiment, which again confirms the BSW prediction. It may be worth emphasizing that in this $t$ domain, the $t$-behavior is definitely not a straight line.

\begin{figure}
 \includegraphics[height=.3\textheight]{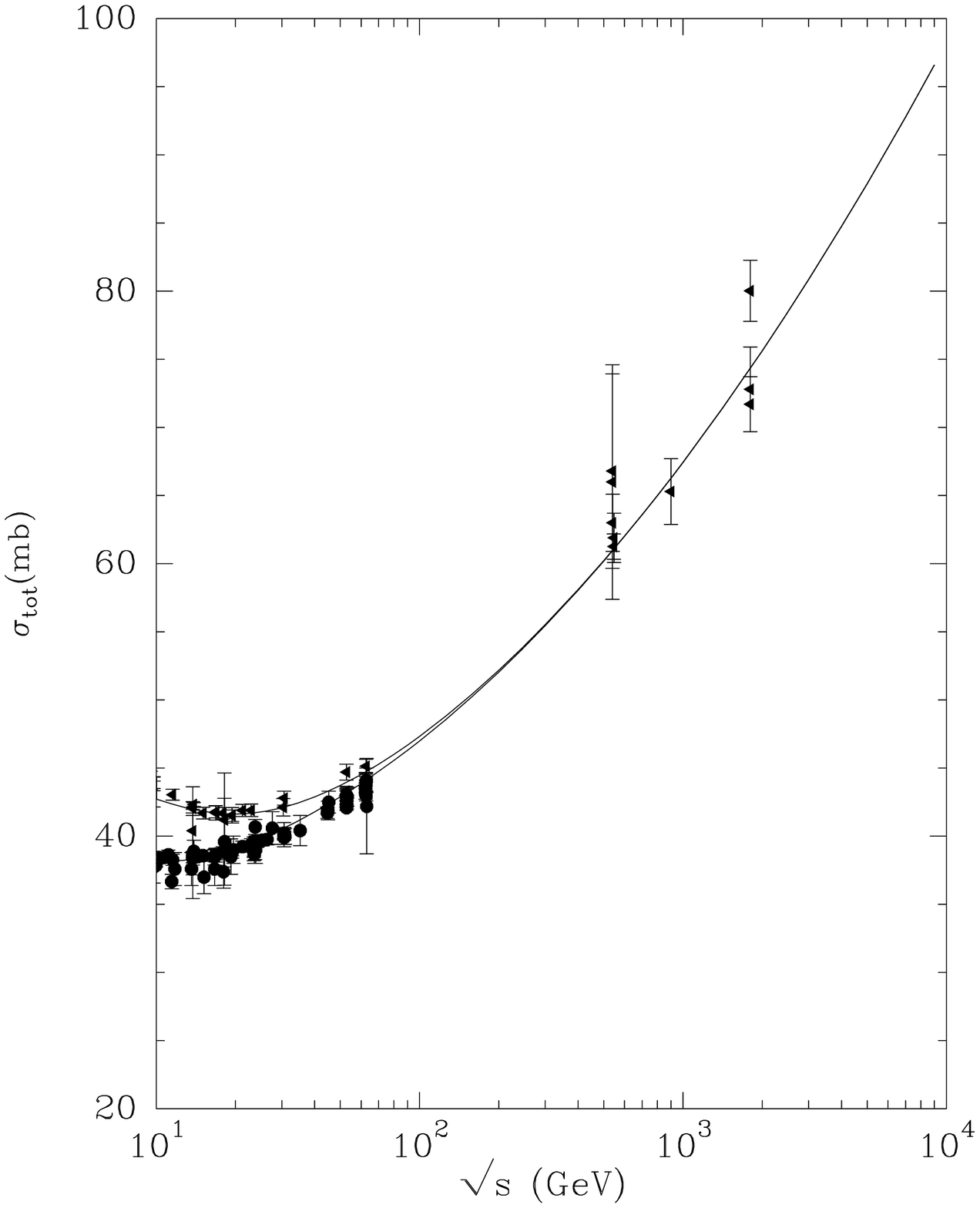}
  \includegraphics[height=.3\textheight]{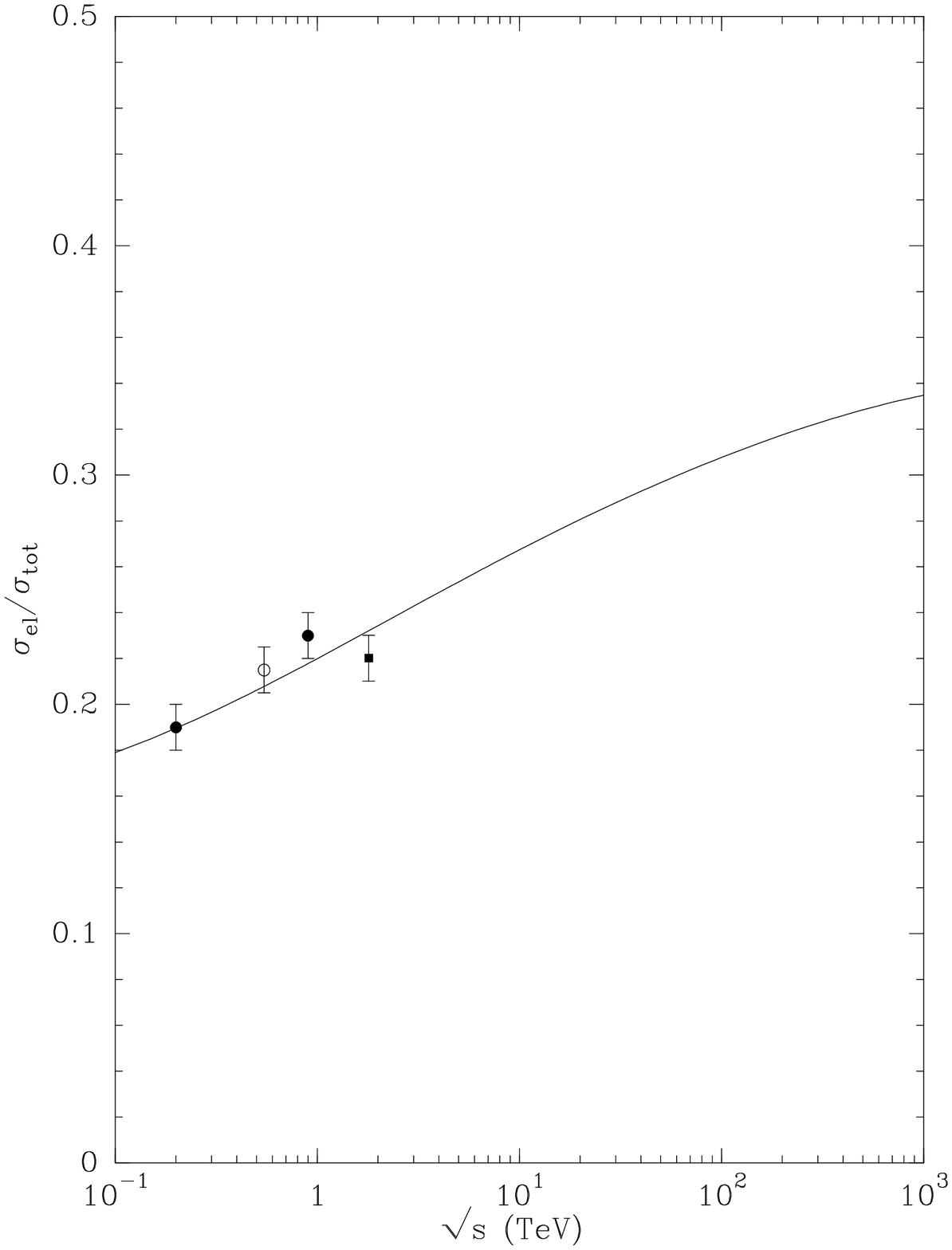}
  \caption{ $pp$ ($\bar pp$) elastic scattering, $\sigma_{tot}$, ({\it Left}), $\sigma_{el}/\sigma_{tot}$ ({\it Right}) as a function of the energy. (Taken from Ref.(4)).}
\label{fig:sig-totel}
\end{figure}

\begin{figure}
 \includegraphics[height=.23\textheight]{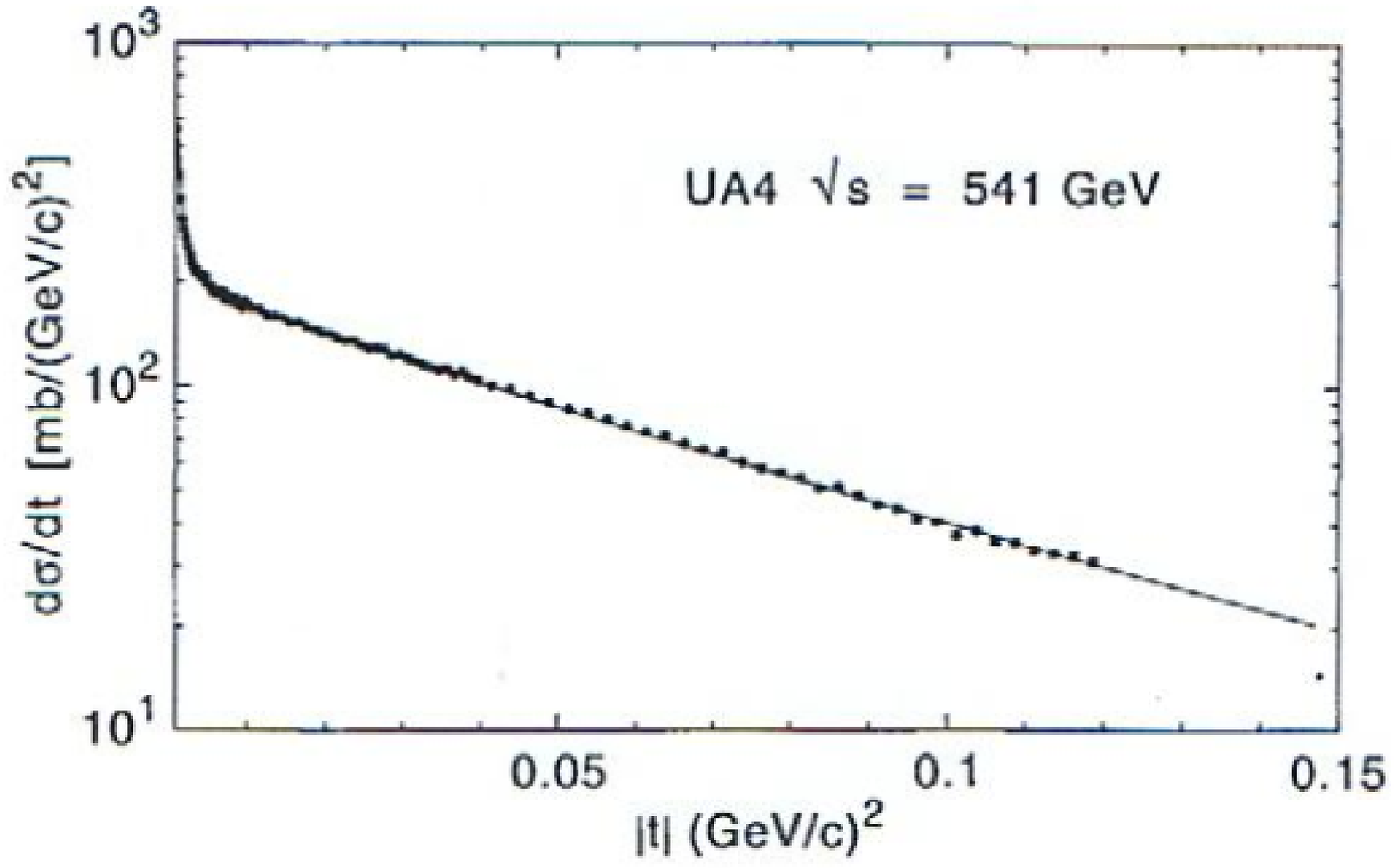}
  \includegraphics[height=.23\textheight]{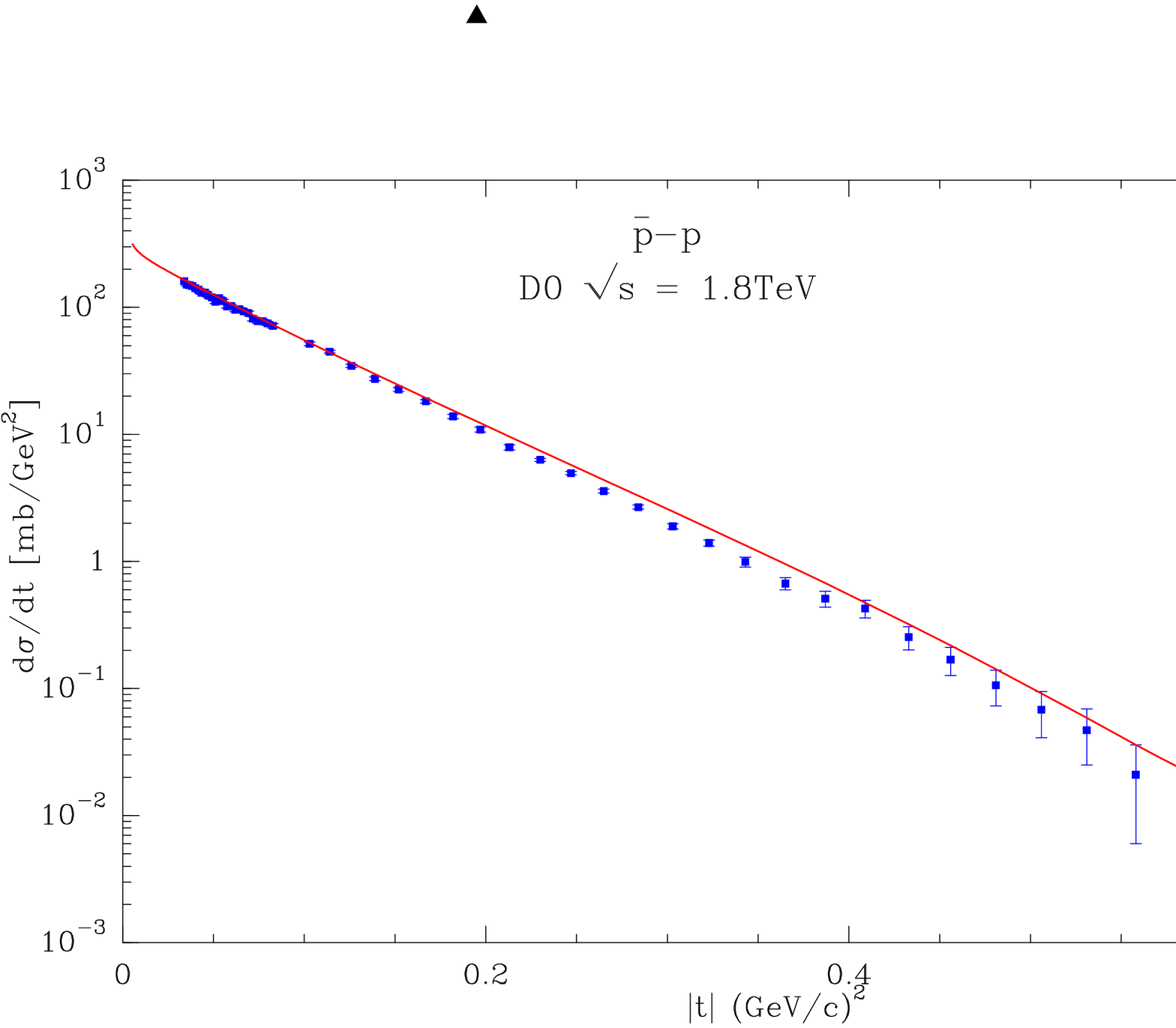}
  \caption{({\it Left}): $d\sigma/dt$ for near-forward $\bar p p$ elastic scattering at $\sqrt{s}$= 541GeV data from Ref. \cite{augier}, the curve is the BSW prediction \cite{bsw87}. ({\it Right}): $d\sigma/dt$ for near-forward $\bar p p$ elastic scattering at  $\sqrt{s}$= 1.8TeV data from Ref. \cite{amos}, the curve is the BSW prediction \cite{bsw2}. }
\end{figure}

\newpage
\section{The LHC energy region}
There are two experiments measuring the proton-proton total cross section:
TOTEM associated with the CMS detector and the ALFA associated with the
ATLAS Collaboration.  The published results for both experiments are for the center-of-mass energy 7 TeV.\\
The BSW approach predicts at 7 TeV $\sigma_{\it tot} = 93.6 \pm 1$mb and $\sigma_{\it el} = 24.8 \pm 0.3$mb.\\
TOTEM has measured at 7 TeV \cite{totem} $\sigma_{\it tot} = 98.0 \pm 2.5$mb and $\sigma_{\it el} = 24.8 \pm 1.2$mb.\\
ATLAS-ALFA has measured at 7 TeV \cite{alfa} $\sigma_{\it tot} = 95.35 \pm 1.36$mb and $\sigma_{\it el} = 24.0 \pm 0.60$mb\\
It is clear that ATLAS-ALFA is more accurate than TOTEM and BSW agrees very well on both results for $\sigma_{\it el}$. However it is below both experimental results on $\sigma_{\it tot}$, although only less than 2 $\sigma$. We look forward to the data after the upgrade of the center-of-mass energy of the LHC next year to 13 TeV.\\
The relevance of the measurement of the  $\rho$ parameter at LHC has been strongly emphasized \cite{bkmsw}, but so far we only have an estimate from TOTEM \cite{diele}, namely $|\rho|= 0.145 \pm 0.091$ with a poor accuracy. There are future plans to reach a much higher precision by measuring the CNI region down
to $|t|$ values of $6 \cdot 10^{-4}\mbox{GeV}^2$.\\
Finally we turn to the differential cross section for near-forward $pp$ elastic scattering and we show in Fig. 3 the available LHC data compared to the BSW prediction, which seems in better agreement with the ATLAS-ALFA data.

\begin{figure}[hp]
 \includegraphics[height=.2\textheight]{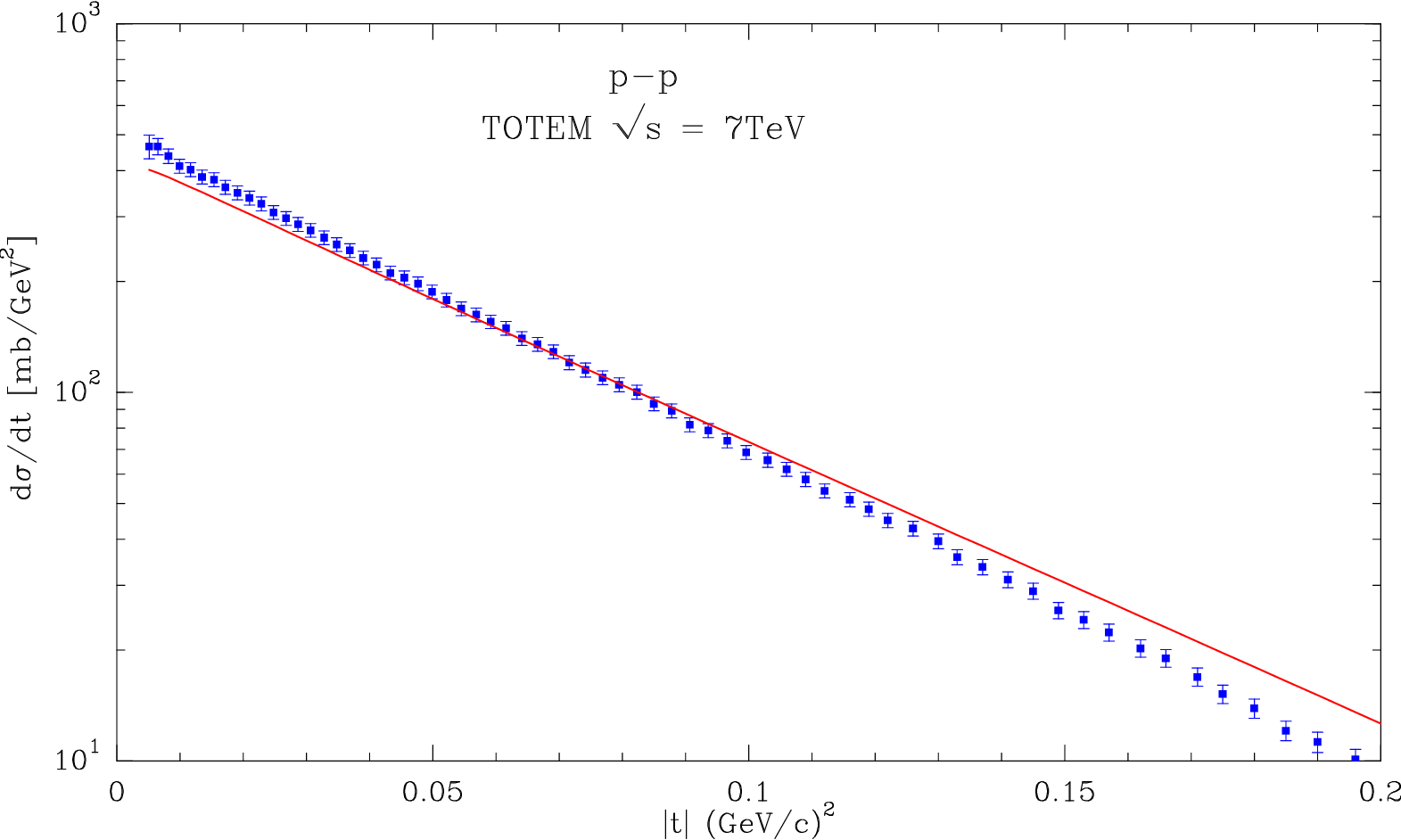}
  \includegraphics[height=.2\textheight]{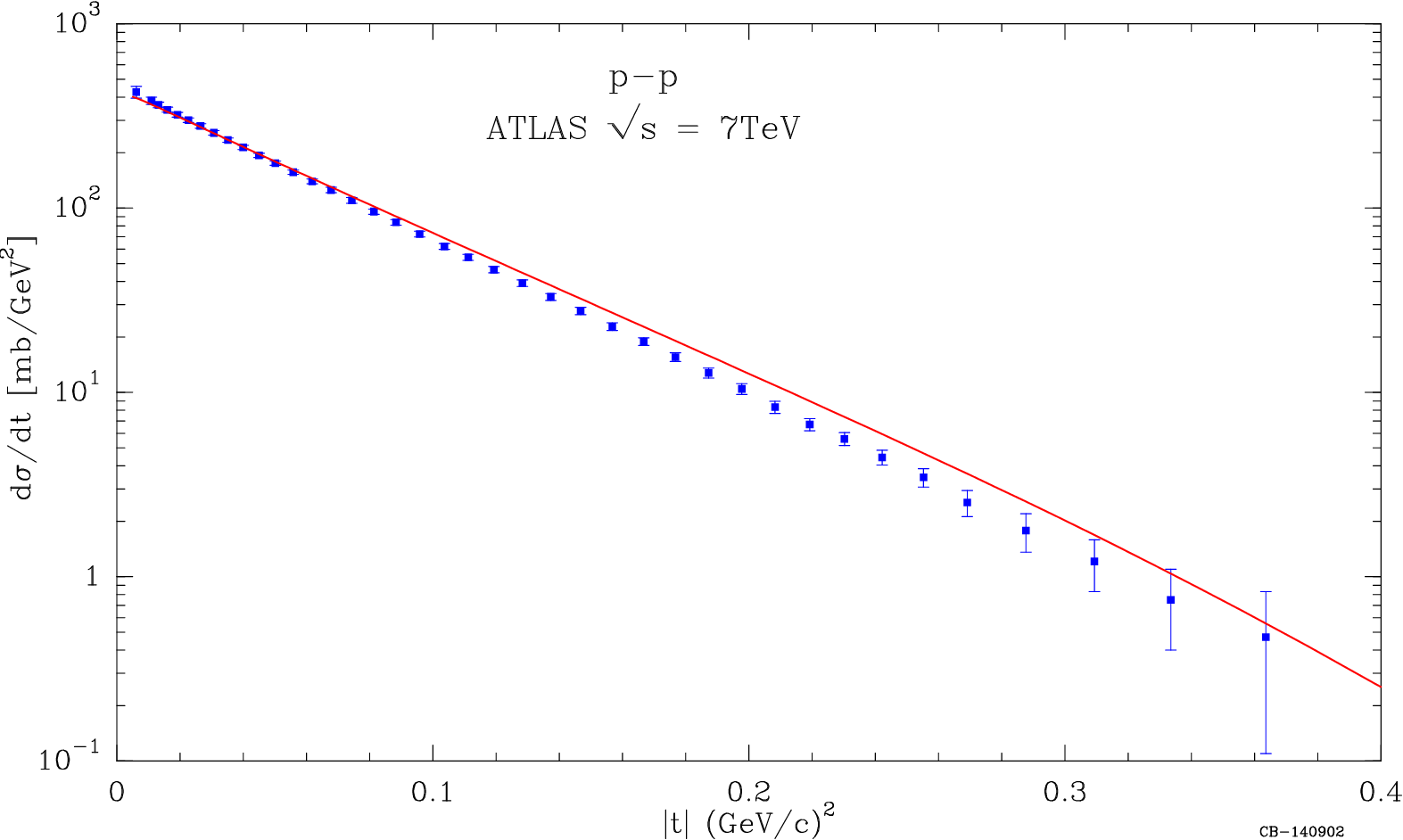}
  \caption{$d\sigma/dt$ for near-forward $pp$ elastic scattering at LHC
and the curve of the BSW prediction. ({\it Left}) TOTEM data \cite{totem}. ({\it Right}) ATLAS-ALFA data \cite{alfa}}
\end{figure}




\bibliographystyle{aipproc}   

\bibliographystyle{aipprocl} 





\end{document}